\documentstyle[preprint,prd,aps,epsfig]{revtex}       
\tightenlines
\begin{document}

\preprint{\vbox{\hbox{\bf IF--1529/2001}}}

%
%

\title{Excited Leptons at the CERN Large Hadron Collider}

\author{O.\ J.\ P.\ \'Eboli$^1$, 
S.\ M.\ Lietti$^1$ and Prakash Mathews$^{2}$  \\ $ $\\}

\address{$^1$Instituto de F\'{\i}sica da USP, \\ 
         C.P. 66.318, S\~ao Paulo, SP 05389-970, Brazil.\\ $ $\\
         $^2$School of Physics, University of Hyderabad, \\
         Hyderabad 500 046, India.}

\maketitle


\begin{abstract}
  
  We analyze the potential of the CERN Large Hadron Collider (LHC) to
  search for excited spin-$\frac{1}{2}$ electrons and
  neutrinos. Assuming an $SU(2)_L \otimes U(1)_Y$ invariant model, we
  study in detail the single production of excited electrons and
  neutrinos and respective backgrounds through the reactions $ p p \to
  e^+ e^- V$ and $e^\pm \nu V$ with $V=\gamma$, $W$, or $Z$.  We show
  that the LHC will be able to tighten considerably the direct
  constraints on these possible new states, probing excited lepton
  masses up to 1--2 TeV depending on their couplings to fermions and
  gauge bosons.

\end{abstract}

\vskip 2pc


\section{Introduction}

The standard model (SM) of electroweak interactions explains very well the
available experimental data \cite{experimental_data}.  Notwithstanding, some
important questions are still left unanswered, in particular, the
proliferation of fermionic generations and their complex pattern of masses and
mixing angles are not explained by the model. A rather natural explanation for
the replication of the fermionic generations is that the known leptons and
quarks are composite \cite{composite} and they should be regarded as the
ground state of a rich spectrum of fermions. Therefore, the observation of
excited quarks and leptons would be an undeniable signal for compositeness
\cite{excited_states}.

\medskip

Up to now all direct searches for compositeness have failed. At 
the DESY $ep$ collider HERA
\cite{hera}, operating in the $e^+ p$ mode, no evidence of excited
fermions was found which leads to 95\% C.L. bounds ${\cal O}(200)$ GeV
on the excited lepton mass for $m^\star = \Lambda$, where we denote by
$\Lambda$ the strong dynamics scale. The direct search for single and
pair productions of excited leptons at the CERN $e^+ e^-$ collider
LEP leads to 95\% C.L. constraints
on $m^\star$ of the order ${\cal O}(100)$ GeV for $m^*=\Lambda$
\cite{lep}.

\medskip

In this work, we reexamine the single production of excited spin-$\frac{1}{2}$
electrons ($e^\star$) and neutrinos ($\nu^\star$) at the LHC via the reactions
\begin{eqnarray}
p p &\to&  e^\pm e^{\star \mp}  \to e^+ e^- V \;\; , 
\label{p1} \\
p p &\to&  \nu e^{\star \pm} + \nu^\star e^\pm \to e^\pm \nu V \;\; , 
\label{p2}
\end{eqnarray}
where $V$ stands for $\gamma$, $W^\pm$ or $Z$.  We carefully analyze
the SM backgrounds and signals for an $SU(2)_L \otimes U(1)_Y$
invariant model, which is described below. We show that the strongest
limits are obtained for $V=\gamma$ and that the CERN Large Hadron
Collider (LHC) will be able to
probe for excited electrons and neutrinos with masses up to 1 TeV,
assuming that $m^*= \Lambda$.

\medskip

The outline of this paper is the following. In section II we review the
model used in our analysis. Section III contains our results while the
conclusions are presented in Section IV.

\section{Model}

The strong dynamics of the lepton constituents is unknown, therefore,
we employ a model-independent analysis of the effects of fermion
compositeness based on effective Lagrangian techniques. In this work,
we assume that the excited fermions have spin and isospin
$\frac{1}{2}$ since the last assignment allows the excited fermions to
acquire their masses prior to $SU(2) \otimes U(1)$ breaking,
avoiding dangerous bounds coming from the precise determinations of
$\Delta \rho$.  In the case of the first generation leptons, the
assumed lightest particle spectrum is
\begin{eqnarray}
l_L= \left[\begin{array}{c} \nu_e  \\ e \end{array}\right]_L 
\;\;\; ; \;\;\;
e_R 
\;\;\; : \;\;\;
L_L= \left[\begin{array}{c} \nu_e^* \\ e^* \end{array}\right]_L
\;\;\; ; \;\;\;
L_R= \left[\begin{array}{c} \nu_e^* \\ e^* \end{array}\right]_R \;\; .
\end{eqnarray}

\medskip 

The Lagrangian describing the transition between ordinary and excited fermions
should exhibit chiral symmetry in order to protect the light leptons from
acquiring radiatively a large anomalous magnetic moment \cite{chiral}. The
$SU(2) \otimes U(1)$ invariant Lagrangian describing the interaction between
excited and ordinary leptons is \cite{effective_Lagrangian}
\begin{eqnarray}
{\cal L}_{l l^*} = \frac{1}{2\Lambda} \bar{L}_R \sigma^{\mu\nu}
\left[g f \frac{\vec{\tau}}{2} \vec{W}_{\mu\nu} + g' f' \frac{Y}{2} 
B_{\mu\nu} \right] l_L + h.c. \;\; ,
\label{lag_1}
\end{eqnarray}
where $\Lambda$ is the compositeness scale while $g$ and $g'$ are,
respectively, the $SU(2)_L$ and $U(1)_Y$ coupling constants. Here the
constants $f$ and $f'$ are weight factors which can be interpreted as
different scales $\Lambda_i = \Lambda/f_i$ for the gauge groups. As
usual, the tensors $\vec{W}_{\mu\nu}$ and $B_{\mu\nu}$ represent the
field-strength tensors. Notice that our hypothesis implies that only the
right-handed part of the excited fermions takes part in the
generalized magnetic interaction with the known leptons.

\medskip

In the physical basis, the Lagrangian (\ref{lag_1}) can be written as
\begin{eqnarray}
{\cal L}_{l l^*} = \frac{e_0}{2\Lambda} (f-f') N_{\mu\nu}
\sum_{l=\nu_e,e} \bar{l^*} \sigma^{\mu\nu} l_L
+ \frac{e_0}{2\Lambda} f \sum_{l,l'=\nu_e,e} \Theta_{\mu\nu}^{\bar{l^*},l}
\bar{l^*} \sigma^{\mu\nu} {l'}_L
+ h.c. \;\; ,
\label{lag_2}
\end{eqnarray} 
where $\ell$ ($\ell^\star$) stands for $\nu_e$ or $e$ ($\nu^*_e$ or $e^*$) and
$e_0$ is the proton electric charge.  The first term contains only triple
vertices with $N_{\mu\nu} = \partial_\mu A_\nu -(s_w/c_w)\partial_\mu Z_\nu$
and it vanishes for $f=f'$. On the other hand, the second term contains triple
as well as quartic vertices with
\begin{eqnarray}
\Theta_{\mu\nu}^{\bar{\nu^*_e},\nu_e} &=& \frac{1}{s_w c_w} \partial_\mu Z_\nu 
- i \frac{e}{s_w^2} W^+_\mu W^-_\nu \;\; ,
\nonumber \\
\Theta_{\mu\nu}^{\bar{e^*},e}&=&-\left(2 \partial_\mu A_\nu +
\frac{c_w^2-s_w^2}{s_w c_w} \partial_\mu Z_\nu - 
i \frac{e}{s_w^2} W^+_\mu W^-_\nu \right) \;\; ,
\nonumber \\
\Theta_{\mu\nu}^{\bar{\nu^*_e},e} &=& \frac{\sqrt{2}}{s_w} \left[
\partial_\mu W^+_\nu - i e W^+_\mu \left(A_\nu+\frac{c_w}{s_w}Z_\nu 
\right) \right]\;\; ,
\\
\Theta_{\mu\nu}^{\bar{e^*},\nu_e}&=& \frac{\sqrt{2}}{s_w} \left[
\partial_\mu W^-_\nu + i e W^-_\mu \left(A_\nu+\frac{c_w}{s_w}Z_\nu 
\right) \right] \;\; .
\nonumber
\end{eqnarray} 
Adding all the contributions, the chiral $V l^* l$ vertex is
\begin{eqnarray}
\Gamma_\mu^{V\bar{f}^*f} = \frac{e_0}{2\Lambda} q^\nu \sigma_{\mu\nu} 
(1-\gamma_5)f_V \;\; ,
\end{eqnarray} 
with $V=W$ , $Z$,or $\gamma$ and $q$ being the incoming $V$
momentum. The weak and electric charges, $f_W$, $f_Z$ , and $f_\gamma$
are
\begin{eqnarray}
f_W &=& \frac{1}{\sqrt{2}s_w} f  \;\; , 
\nonumber \\
f_Z &=& \frac{4 I_{3L}(c_w^2f+s_w^2f')-4e_f s_w^2f'}{4s_w c_w} \;\; ,
\\
f_\gamma &=& e_f f'+I_{3L}(f-f') \;\; ,
\nonumber 
\end{eqnarray} 
where $e_f$ is the excited fermion charge in units of the proton
charge, $I_{3L}$ is its weak isospin, and $s_w~(c_w)$ is the sine
(cosine) of the weak mixing angle.

\medskip

We present our results using the two complementary coupling assignments 
$f=f'$ and $f=-f'$. For example, for the case $f=f'$ ($f=-f'$), the 
coupling of the photon to excited neutrinos (electrons) vanishes.
In order to illustrate the changes in the phenomenology when we
vary $f$ and $f^\prime$ we display in Table \ref{br} the branching
ratios for excited electrons and neutrinos for the above choices of
couplings.

\section{Signals at Large Hadron Colliders}

In this paper we analyze the potentiality of the LHC to directly
search for excited electrons and neutrinos via the reactions
(\ref{p1}) and (\ref{p2}).  The signal and backgrounds were simulated
at the parton level with full tree level matrix elements, taking into
account interference effects between the SM and excited lepton
contributions. This was accomplished by numerically evaluating
helicity amplitudes for all subprocesses using MADGRAPH
\cite{MadGraph} in the framework of HELAS \cite{helas}, with the new
interactions being implemented as additional Fortran routines.  In our
calculations we used the Martin--Roberts--Stirling Set G
[MRS (G)] \cite{mrs} proton structure functions
with the factorization scale $Q^2 = \hat{s}$.

\medskip

Let us, initially, concentrate on the case in which 
the produced vector boson is a
$\gamma$, that is, the reactions $pp \to e^+ e^- \gamma$ and $pp \to e^\pm \nu
\gamma$. At tree level, these final states can be obtained through the single
production of excited leptons via the Drell--Yan mechanism, followed by their
radiative decays, {\it i.e.}
\begin{eqnarray}
&&q \bar{q} \to Z^\star/\gamma^\star \to e^\pm {e^*}^\mp \to e^\pm e^\mp
\gamma \;\; ,
\\
&&q q' \to W^{\star\pm} \to \nu {e^*}^\pm +  e^\pm \nu^* \to  \nu e^\pm 
\gamma \;\; .
\end{eqnarray}

\medskip

We applied to the above processes the following acceptance cuts
\begin{eqnarray}
&&p_T > 20 \text{ GeV} \;\; ,
\nonumber \\
&&|\eta_{e^\pm,\gamma}| < 2.5 \;\; ,
\label{acceptance} \\
&&\Delta R_{(e^+e^-),(e^+\gamma),(e^-\gamma)} > 0.4 \;\; ,
\nonumber
\end{eqnarray} 
where $p_T$ is the transverse momentum of the visible particle or the
missing transverse momentum when a neutrino is produced. $\eta$ stands
for the pseudo-rapidity of the visible particles and $\Delta R$ ($
=\sqrt{ \Delta \eta^2 + \Delta \phi^2}$) is the separation between two
of them. After applying these initial cuts, the SM cross sections are
\begin{eqnarray}
&&\sigma_{pp \to e^+e^-\gamma} = 1.29 \text{ pb} \;\; , 
\nonumber \\
&&\sigma_{pp \to e^\pm \nu \gamma} = 2.88 \text{ pb} \;\; .
\nonumber
\end{eqnarray} 

\medskip

\begin{figure}
\vskip 50pt
\begin{center}
\parbox[c]{3.5in}{
\mbox{\epsfig{file=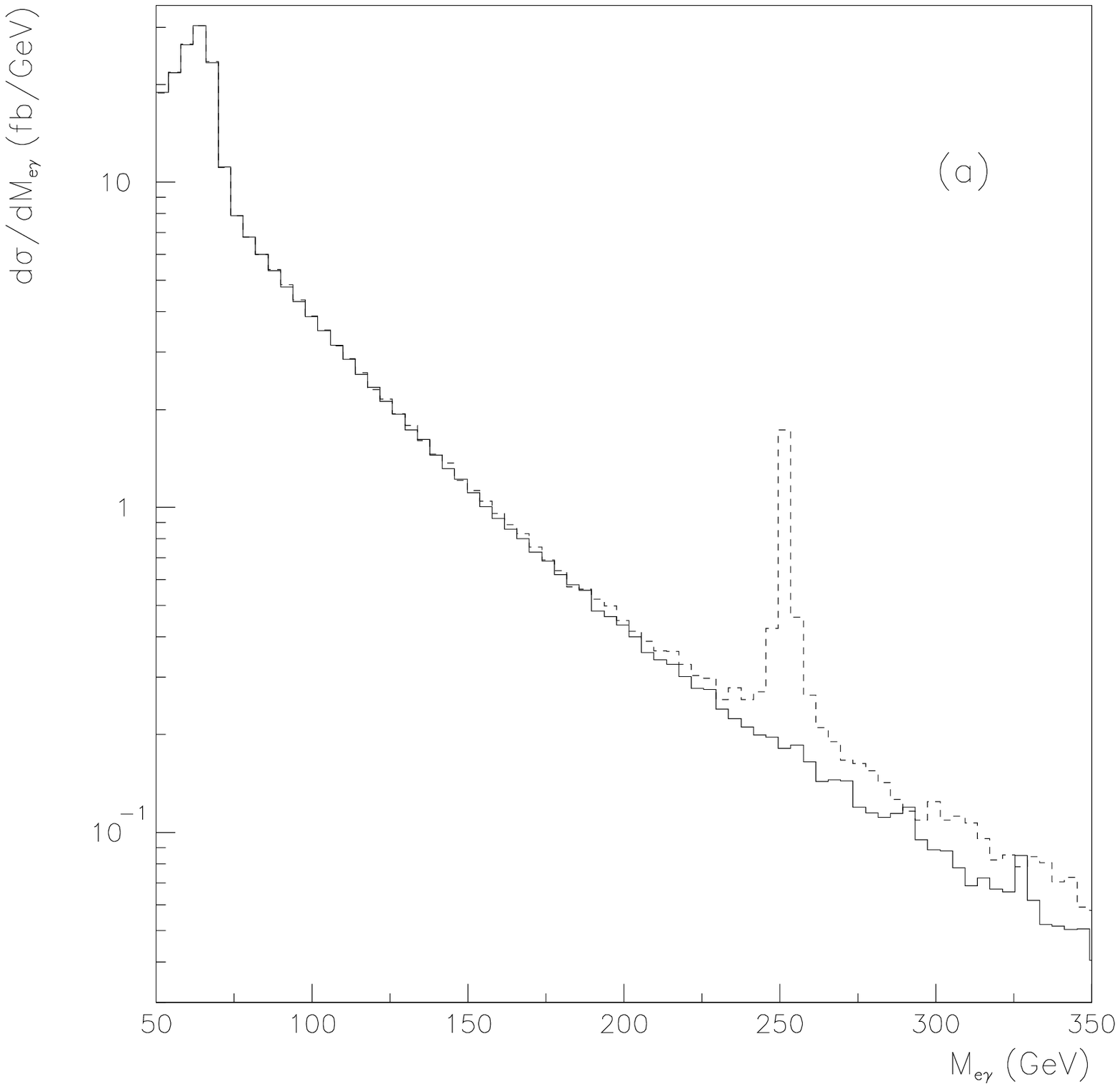,width=\linewidth}}
  }
\parbox[c]{3.5in}{
\mbox{\epsfig{file=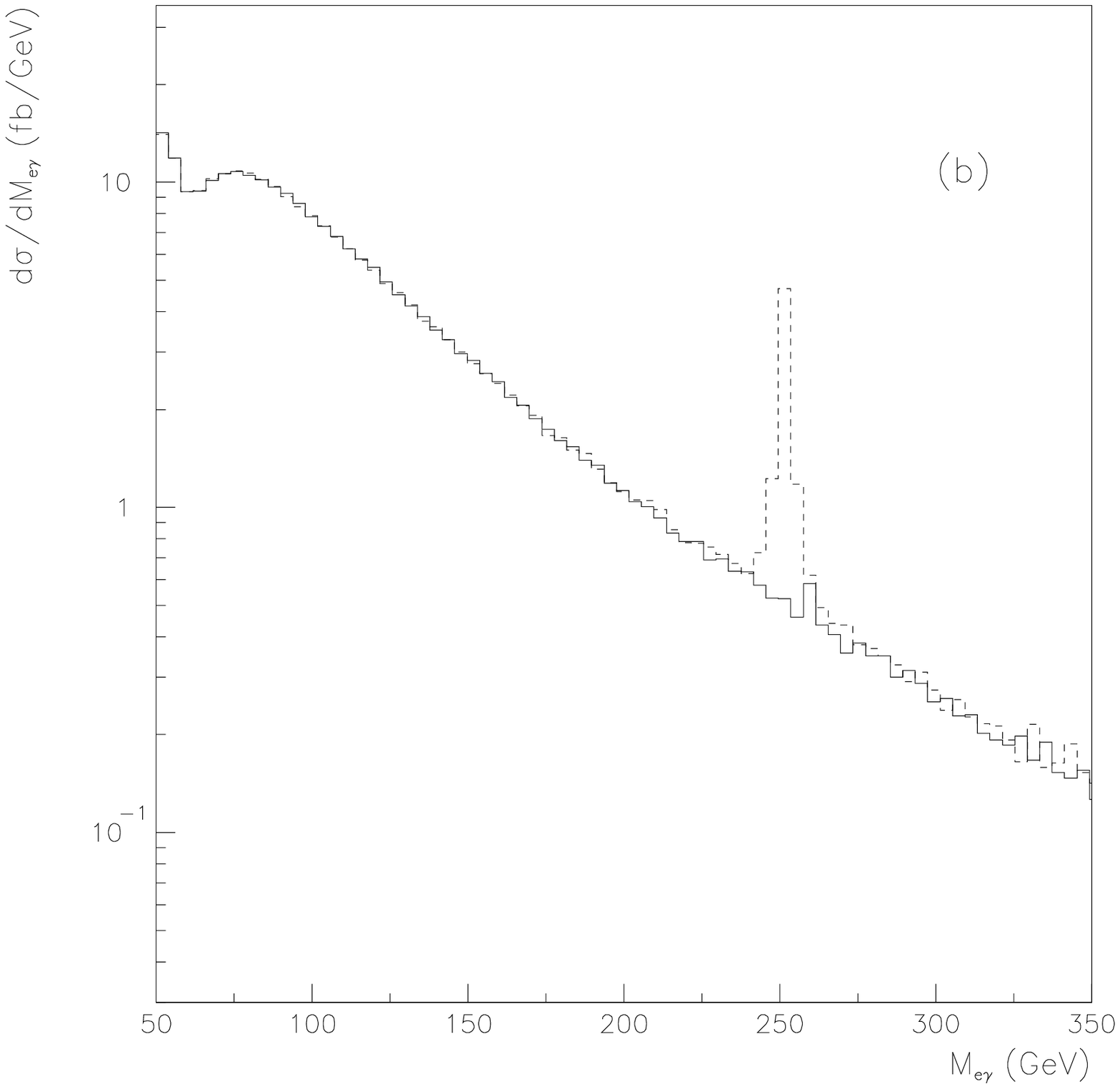,width=\linewidth}}
}
\end{center}
\vskip -50pt
\caption{
        Invariant mass distribution of the pair $e \gamma$ in the
        process $pp \to e^+ e^- \gamma$ (a) and $pp \to e^\pm \nu
        \gamma$ (b).  The full line stands the SM background and the
        dashed line for the excited electron signal, assuming $m^\star
        = 250$ GeV and $f/\Lambda=f^\prime/\Lambda=5/$TeV.}
\label{distributions_eea}
\end{figure}

A natural way to extract the excited electron signal, and at the same time
suppress the SM backgrounds, is to impose a cut on the $e\gamma$ invariant
mass.  For instance, Fig.\ \ref{distributions_eea}(a) [(b)] 
contains the $e\gamma$
invariant distribution in the reaction $pp \to e^+e^- \gamma$ [$e^\pm \nu
\gamma$] for the SM and with the inclusion of an excited electron with mass
$m^\star = 250$ GeV and $f/ \Lambda=f^\prime/\Lambda = 5$ TeV$^{-1}$.  From
these figures we can see that it is convenient to collect the data in
$e\gamma$ invariant mass bins since the signal is concentrated in a small
region of this invariant mass spectrum. Therefore, we introduced the cut
\begin{equation}
|M_{e^\pm \gamma} - \overline{M}| < 25 \text{ GeV} \;\; ,
\label{inv_mass_eg} 
\end{equation}
where $\overline{M}$ is the center of a 50 GeV invariant mass bin. Certainly
this cut is ideal for excited electrons of mass $m^\star = \overline{M}$.
However, the cut (\ref{inv_mass_eg}) is efficient only for excited electron
masses up to 1250 GeV since heavier excited electrons are rather broad
resonances and we also run out of statistics.  Consequently, we only
considered $\overline{M} \le 1500$ GeV. To search for heavier excited
electrons ($m^\star > 1500$ GeV), we performed a new analysis replacing the
cut (\ref{inv_mass_eg}) by
\begin{equation}
       M_{e^\pm \gamma} > 1250 \text{ GeV} \;\; .
\label{inv_mass_eg_hm} 
\end{equation}

To further reduce the SM background, we also vetoed events exhibiting $Z$'s
decaying into $e^+e^-$ pairs through the cut
\begin{equation}
M_{e^+ e^-} > 120 \text{ GeV}.
\label{inv_mass_ee}
\end{equation}

\medskip

\begin{figure}
\protect
\vskip  80pt
\centerline{\mbox{\epsfig{file=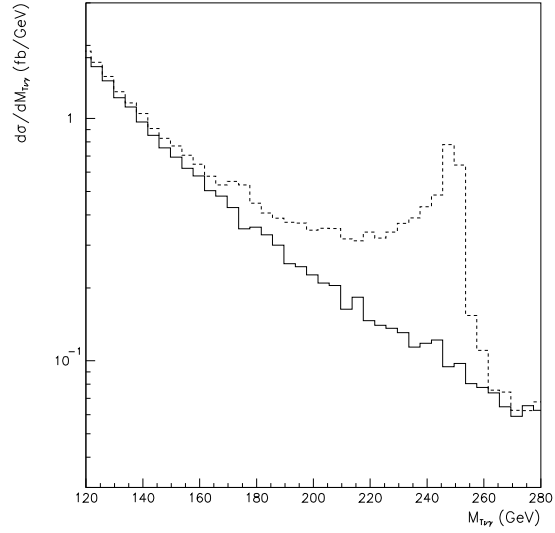,angle=0,width=0.5\textwidth}}}
\vskip -60pt
\caption{
        Transverse mass distribution in the reaction $pp \to e \nu
        \gamma$ at LHC. The full line represents the SM background
        while the dashed line stands for an excited neutrino signal,
        assuming $m^\star = 250$ GeV and $f/ \Lambda = f^\prime/
        \Lambda=5/$TeV.}
\label{distributions_eva}
\end{figure}

Excited neutrinos contribute only to the $e^\pm \nu \gamma$ production
and they can be identified by the $\gamma /\!\!\!p_T$ transverse mass
($M_{T}$) distribution; see Fig.\
\ref{distributions_eva}. Analogously to the excited electron case, we looked
for an excess of events in $M_T$ bins of 50 GeV via the additional cut
\begin{equation}
|M_{T} + 15\text{ GeV} - \overline{M}|  < 25\text{ GeV } \;\; ,
\label{trans_mass_ng} 
\end{equation} 
where $\overline{M}$ is a variable parameter. This cut enhances the signal of
excited neutrinos whose mass is $\overline{M}$.  Once again, the cut
(\ref{trans_mass_ng}) is efficient only for excited neutrino masses up to 1250
GeV so we restricted $\overline{M} \le 1500$ GeV. For higher masses ($m^\star
> 1500$ GeV), the decay width of the excited neutrino is so large that we
performed a different study replacing the cut (\ref{trans_mass_ng}) by
\begin{equation}
M_{T} > 1250 \text{ GeV} \;\; .
\label{trans_mass_ng_hm} 
\end{equation}

\medskip
 
The above cuts (\ref{inv_mass_eg})--(\ref{trans_mass_ng_hm}) reduce the SM
background drastically. For instance, assuming an excited lepton mass of 250
GeV ($= \overline{M}$) the SM background for the processes (\ref{p1}) and
(\ref{p2}) is reduced to
\begin{eqnarray}
&&\sigma_{pp \to e^+e^-\gamma}^{M_{e\gamma}} = 3.55 \text{ fb} 
\;\; , \nonumber \\
&& \sigma_{pp \to e^\pm \nu \gamma}^{M_{e\gamma} } = 
51.7 \text{ fb} 
\;\; , \nonumber\\
&&\sigma_{pp \to e^\pm \nu \gamma}^{M_T}  
= 12.3 \text{ fb} 
\;\; ,
\nonumber
\end{eqnarray}
where we applied cuts (\ref{acceptance}-\ref{inv_mass_ee}) to the first
two results and (\ref{acceptance}), (\ref{trans_mass_ng})
and (\ref{trans_mass_ng_hm}) to the last one.  Tables
\ref{back1}--\ref{back3} contain the cross sections of the
irreducible background after cuts for several excited lepton masses
where we can see that the background diminishes very fast with the
increase of $m^\star$.

\medskip

At this point, it is important to consider other possible sources of
background to these final states since the irreducible background has been
largely reduced.  For example, additional backgrounds are $pp \to e^+e^-
\text{ jet}$ and $pp \to e^\pm \nu \text{ jet}$, where a jet is
misidentified as a photon. Taking the jet faking photon probability at the LHC
to be $f_{\text{fake}} = 1/5000$ \cite{pedro}, we present also in Tables
\ref{back1}--\ref{back3} the expected cross section for these processes.
Another possible reducible background for the process $pp \to e^\pm \nu
\gamma$ is the reaction $pp \to e^+ e^- \gamma$ with one of the charged
leptons escaping undetected, that leads to missing transverse
momentum.  We also evaluated this processes, however, its cross
section turns out to be negligible.

\medskip

In order to quantify the LHC potential to search for excited leptons, we
defined the statistical significance $S$ of the signal
\begin{equation}
S = \frac{|\sigma_{\text{total}} -
\sigma_{\text{back}}|}{\sqrt{\sigma_{\text{back}}}} \;
\sqrt{{\cal L}} \;\; ,
\label{sig}
\end{equation}
where ${\cal L}$ is the LHC integrated luminosity, that we assumed to
be 100 fb$^{-1}$. $S$ can be easily evaluated using Tables
\ref{back1}--\ref{back3} and the expected signal cross section. In order to
derived the attainable limits at the LHC, we assumed that the observed number
of events is the one predicted by the SM.

\medskip

Let us start our analysis by the search for excited electrons. In this case,
we assumed that $f=f^\prime$ in order to reduce the number of free parameters,
and we imposed the cuts (\ref{acceptance}) -- (\ref{inv_mass_ee}) with
$\overline{M}=m^\star$ to maximize the sensitivity for excited fermions of
mass $m^\star$.  We display in Figure \ref{limits}(a) the 95\% C.L. 
bounds on the
coupling $|f/\Lambda|$, coming from the process $ pp \to e^+ e^- \gamma$
($e^\pm \nu \gamma$), as a function of the excited electron mass. As we can
see, the $e^+ e^- \gamma$ production leads to slightly better limits on
excited electrons except at small $m^\star$. The combined results of these two
processes are also presented in this figure and these bounds turn out to be at
least an order of magnitude more stringent than the present best limits coming
from the HERA experiments.  Moreover, the LHC will be able to extend
considerably the range of excited electron masses that can be probed (up to 2
TeV).

\medskip

In the study of the excited neutrino production, we assumed
$f=-f^\prime$ that leads to a non-vanishing $\nu_e \nu_e^\star \gamma$
coupling. The 95\% C.L. limits on $|f/\Lambda|$ that can be obtained
from the reaction $p p \to e^\pm \nu \gamma$, where we imposed the
cuts (\ref{acceptance}) and (\ref{trans_mass_ng}) --
(\ref{trans_mass_ng_hm}), are shown in Figure \ref{limits}(b).  Notice
that the limits on excited neutrinos from this process are looser
(stronger) than the ones derived for excited electrons for $m^\star <
1400$ ($> 1400$) GeV. Furthermore, these bounds are orders of
magnitude more stringent than the presently available ones and they
span a much larger range of excited neutrino masses.

\medskip

It is also interesting to obtain a bound on the excited electron mass
assuming that $f=f^\prime$ and $f/\Lambda = 1/ m^\star$. In this
scenario, the LHC will be able to rule out excited electrons with
masses smaller than 1 TeV, at 95\% C.L., through the study of either the
$e^+e^-\gamma$ or $e^\pm \nu \gamma$ productions, while their combined
results increase this limit by 40 GeV.  This limit improves the
present HERA bound (233 GeV) \cite{hera} by a factor of roughly
5. In the case of excited neutrinos, assuming $f =- f^\prime$ and
$f/\Lambda = 1 / m^\star$, the analysis of the $e^\pm \nu \gamma$
final state leads to $m^\star > 838$ GeV at 95\% C.L.. Again, this limit
is much more restrictive than the available HERA one ($\sim 150$ GeV).

\medskip

Note that for the choice $f=f^\prime$ ($f=-f^\prime$), the coupling of
excited neutrinos (electrons) to photons vanishes. In this way, the
production of a pair of leptons $ee$ or $e\nu$ with a photon can probe
only the excited electron (neutrino) production if $f=f^\prime$
($f=-f^\prime$). Therefore, we should also consider the excited lepton
decay into a $W$ or a $Z$ and an ordinary lepton in these
cases. Taking into account only the leptonic decay of the weak gauge
bosons, we also analyzed the processes $pp \to e^+ e^- e^+ e^-$ and
$e^+ e^- e^\pm \nu$. Notice that excited electrons can contribute to
both reactions, while excited neutrinos only to the second one.

\medskip

In order to look for excited electrons in the $e^+ e^- e^+ e^-$ and
$e^+ e^- e^\pm \nu$ productions, we applied initially the acceptance
cut (\ref{acceptance}) and then required
\begin{equation}
M_{e^+e^-} > 20 \hbox{ GeV} 
\label{cut_photon}
\end{equation}
for all possible $e^+e^-$ pairs. This cut reduces the SM contribution
due to photon exchange. The SM background to the $e^+ e^- e^+ e^-$
production receives a large contribution from the $Z$ pair production
and it can be further suppressed by vetoing events that exhibit two
$e^+e^-$ pairs compatible with being a  $Z$, {\em i.e.}
\begin{eqnarray}
| M_{e^+_1e^-_1} - m_Z| < 25 \hbox{ GeV} &&\hbox{ and } 
| M_{e^+_2e^-_2} - m_Z| < 25 \hbox{ GeV}
\nonumber \\
&&\hbox{ or}
\label{cut_z}
\\
| M_{e^+_1e^-_2} - m_Z| < 25 \hbox{ GeV} &&\hbox{ and }
| M_{e^+_1e^-_2} - m_Z| < 25 \hbox{ GeV,}
\nonumber
\end{eqnarray}
where $e_1^\pm$ ($e^\pm_2$) is the electron/positron with the highest
(smallest) energy.

\medskip

The SM background for the excited neutrino search in the $e^+ e^-
e^\pm \nu$ channel can be depleted by requiring that the transverse
mass calculated using all charged leptons satisfy
\begin{equation}
          M_{T_{e\nu}} > 20 \hbox{ GeV.} 
\label{cut_mtphoton}
\end{equation}
Another important SM contribution to this reaction is $WZ$ production
with the $Z$ decaying into a pair $e^+ e^-$ and the $W$ decaying into
a pair $e\nu$. In order to reject this process, we vetoed events
displaying a pair $e^+e^-$ compatible with being a $Z$ and the invariant
mass of the remaining $e^\pm$ close to the $W$ mass, {\em i.e.}
\begin{equation}
| M_{e^\pm e^\mp} - m_Z| < 25 \hbox{ GeV} \hbox{ and }
| M_{T_{e^\pm \nu}} - m_W + 15\hbox{ GeV }| < 25 \hbox{ GeV.} 
\label{cut_w}
\end{equation} 
Moreover, the excited electron signal in this topology originates from its
charged current production in association with a neutrino.  Therefore, the
three charged leptons in the final state come from the decays of the excited
electron. To further enhance the signal we performed the analysis considering
bins in the invariant mass of the three charged leptons in the final state
$M_{eee}$, {\em i.e.} we demanded that
\begin{equation}
|M_{eee}-\overline{M}| < 25 \hbox{ GeV,} 
\label{cut_eee}
\end{equation} 
where the center of the bin $\overline{M}$ is again a variable parameter.

\medskip

We present in Figure \ref{limits_4}(a) the attainable 95\% C.L. limits on
excited electrons coming from the $e^+e^-e^+e^-$ and $e^+ e^- e^\pm
\nu$ reactions, assuming that $f/\Lambda = f^\prime /\Lambda$ and
$f/\Lambda = -f^\prime/\Lambda$.  As we can see, the bounds for
$f/\Lambda = f^\prime /\Lambda$ are an order of magnitude weaker than
the ones originating from the decay of the excited electron into a
photon--electron pair.  Notwithstanding, these processes are important
when this decay channel is closed. As expected, the four lepton bounds
for $f/\Lambda = -f^\prime /\Lambda$ are ${\cal O}(20)$\% more
restrictive than the ones for $f/\Lambda = f^\prime/\Lambda$.

\medskip  

The charged current production of excited neutrinos can contribute
only to the $e^+ e^- e^\pm \nu$ final state. The decay of the excited
neutrino either in $e W$ or in $\nu Z$ leads to $e^+ e^- \nu$ once we
consider the $W$ and $Z$ decay into first family leptons. Therefore,
the transverse mass of the $e^+e^-$
\begin{equation}
M_{T_{ee\nu}} = \sqrt{2(p_{T_{ee}}   p_{T_{miss}}
              - \vec{p}_{T_{ee}} \cdot \vec{p}_{T_{miss}})}\;\;\;,
\end{equation}
where $p_{ee} = p_{e^+} + p_{e^-}$ and $p_{T_{ee}}$ is its transverse momentum,
characterizes the excited neutrino production. In order to isolate the
excited neutrino signal in the $e^+ e^- e^\pm \nu$ topology, we
initially applied the cuts (\ref{acceptance}), (\ref{cut_photon}), and
(\ref{cut_mtphoton}) and then we required the event to present an $e^+
e^-$ pair with a transverse mass in the bin
\begin{equation}
|M_{T_{ee\nu}}+ 15\text{ GeV} - \overline{M}|  < 25\text{ GeV,} 
\label{cut_eenu}
\end{equation}
with $\overline{M}$ being a variable parameter that enhances the search for
excited neutrinos with mass $m^\star = \overline{M}$.

\medskip

We present in Figure \ref{limits_4}(b) the attainable 95\% C.L. limits on
excited neutrinos coming from the $e^+ e^- e^\pm \nu$ reaction,
assuming that $f/\Lambda = f^\prime /\Lambda$ and $f/\Lambda = -
f^\prime /\Lambda$. As we can see, these bounds are an order of
magnitude weaker than the ones coming from the decay of the excited
neutrino into a photon-neutrino pair for $f = - f^\prime$.

\section{Summary and Conclusions}

\begin{figure}
\begin{center}
\parbox[c]{3.5in}{
\mbox{\epsfig{file=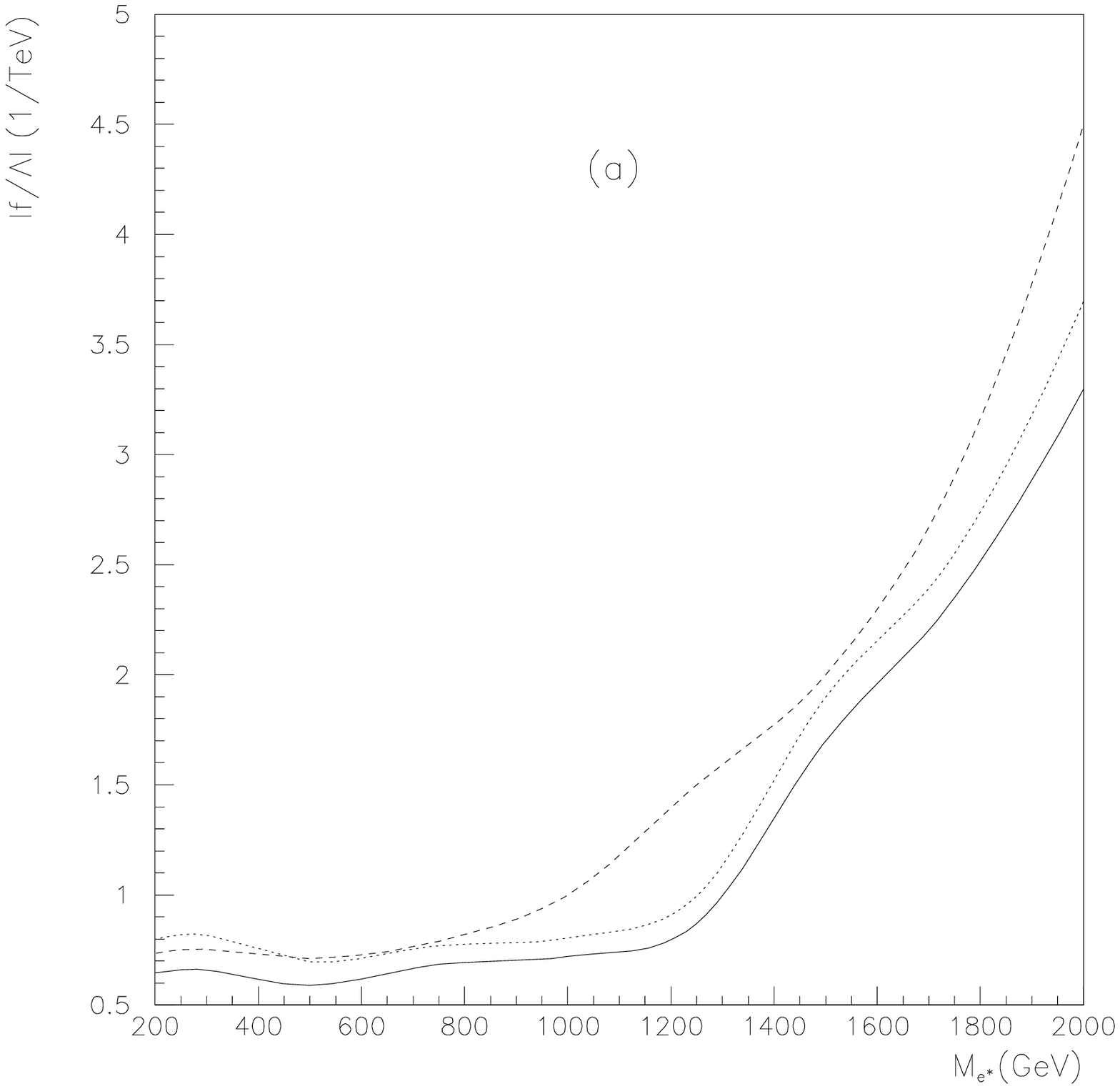,width=\linewidth}}
  }
\parbox[c]{3.5in}{
\mbox{\epsfig{file=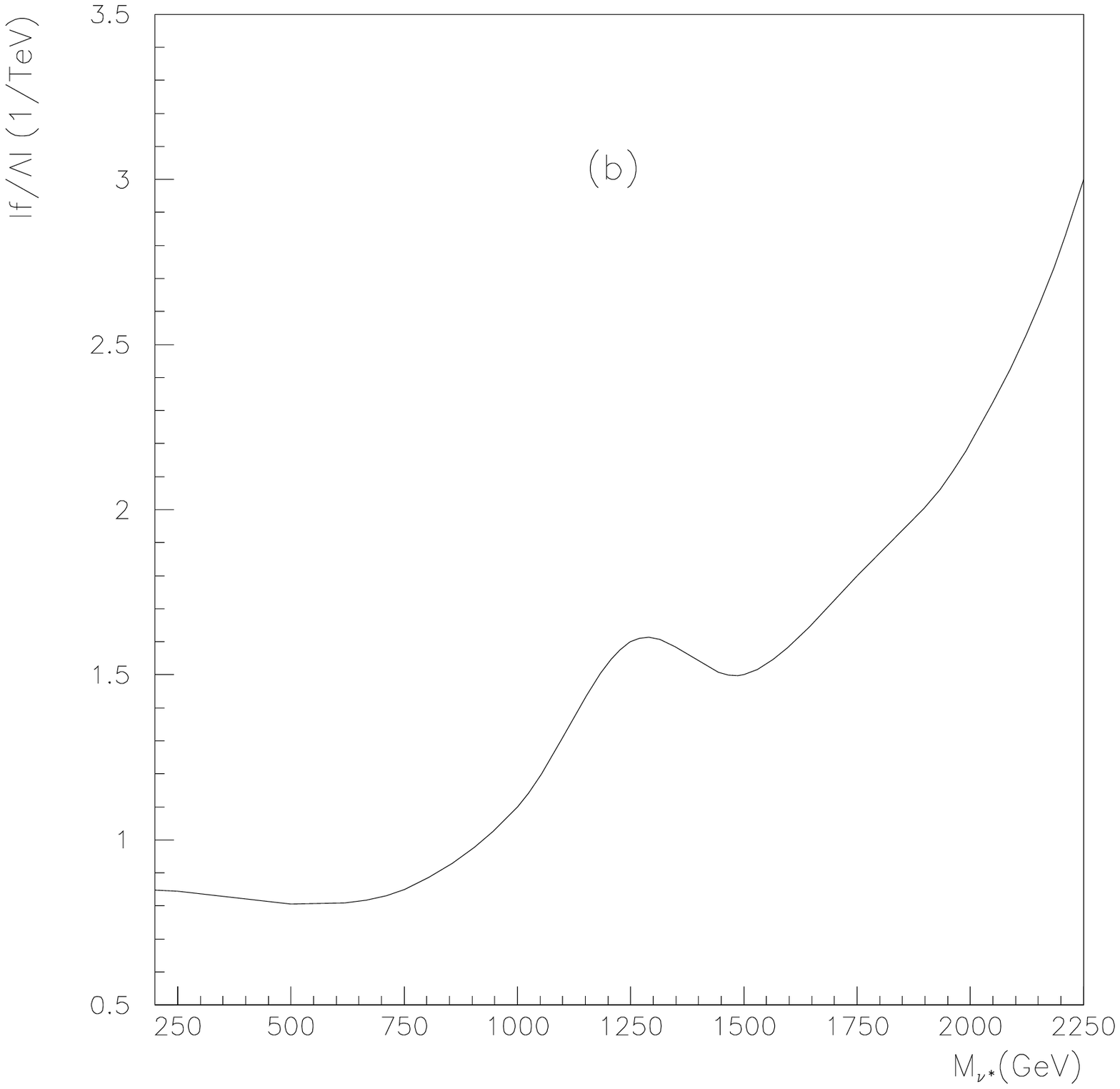,width=\linewidth}}
}
\end{center}
\vskip -75pt
\caption{
        95\% C.L. limits on the coupling $|f /\Lambda|$ of excited
        electrons (a) and excited neutrinos (b). In (a) the dotted
        (dashed) line stands for the bounds coming from the $pp \to
        e^\pm \nu \gamma$ ($ e^+ e^- \gamma$) reaction while the solid
        line represents the combined results. In (b), the solid line
        displays the bounds coming from the process $pp \to e^\pm \nu
        \gamma$.  }
\label{limits}
\end{figure}

\begin{figure}
\vskip  80pt
\begin{center}
\parbox[c]{3.5in}{
\mbox{\epsfig{file=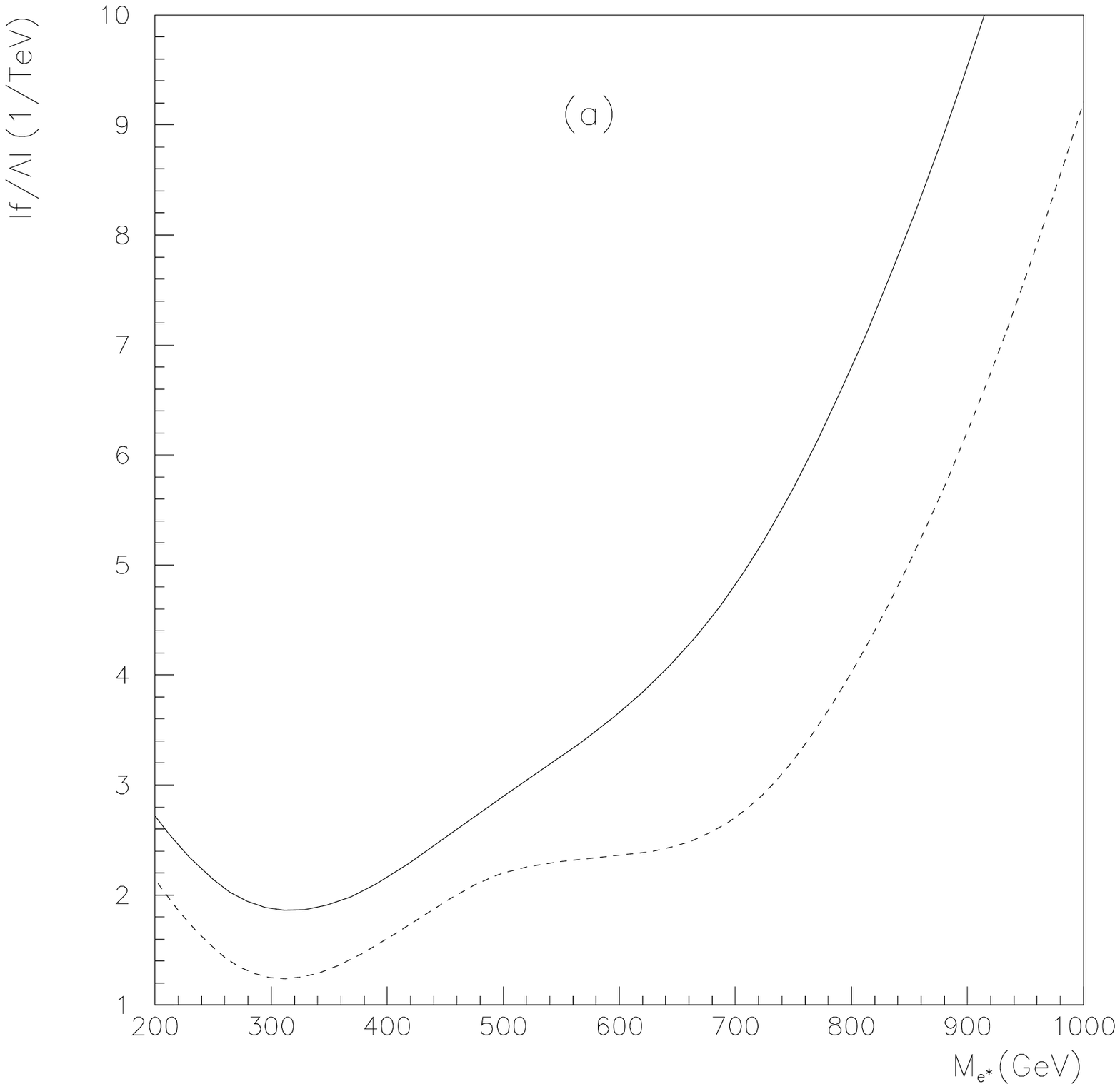,width=\linewidth}}
  }
\parbox[c]{3.5in}{
\mbox{\epsfig{file=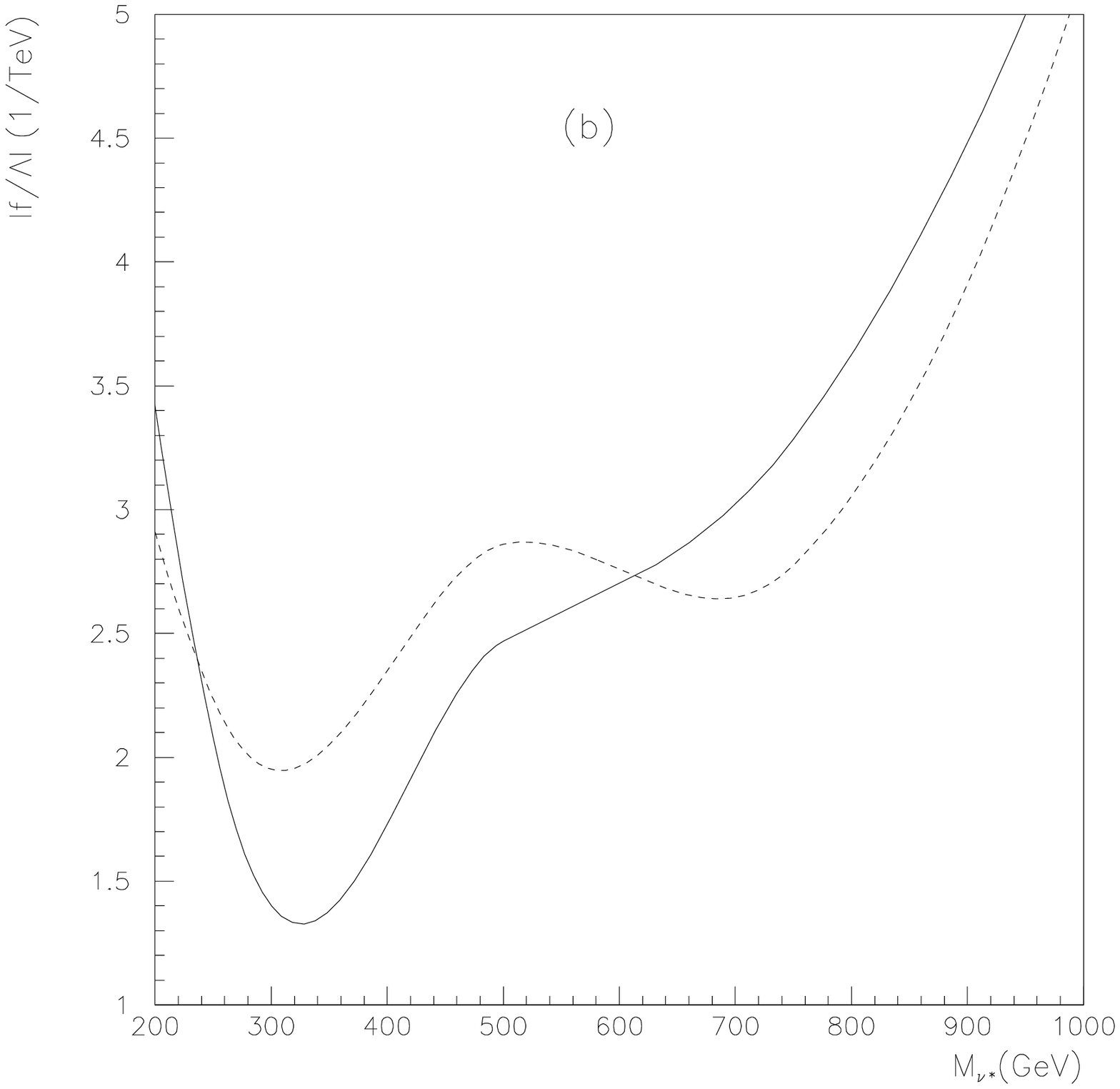,width=\linewidth}}
}
\end{center}
\vskip -75pt
\caption{
        95\% C.L. limits on the coupling $|f /\Lambda|$ of excited
        electrons (a) and excited neutrinos (b) obtained by combining
        the searches in the reactions $pp \to e^+ e^- e^\pm \nu$ and
        $pp \to e^+ e^- e^+ e^-$.  The solid (dashed) line stands for
        the coupling assignment $f=f'$ ($f=-f'$).  }
\label{limits_4}
\end{figure}

We analyzed the potential of the LHC to unravel the existence of
excited leptons through the study of the processes (\ref{p1}) and
(\ref{p2}). We assumed a center-of-mass energy of 14 TeV and an integrated 
luminosity of 100 fb$^{-1}$ in our calculations. The final states 
containing a photon ($e^+e^-\gamma$ or $e^\pm \nu \gamma$) 
lead to the most stringent bounds, as can be seen in Figure (\ref{limits}), 
provided the excited lepton ($\ell^\star$) has a sizable branching ratio 
into pairs $\ell \gamma$. Otherwise, the search for excited leptons should 
be carried out studying the final states $e^+ e^- e^+ e^-$ and 
$e^+ e^- e^\pm \nu$ where our results are presented in Figure 
(\ref{limits_4}). We also considered the possibility of the $V$ boson 
in the processes (\ref{p1}) and (\ref{p2}) decaying into muons, however the 
improvement in the bounds is marginal. 

\medskip 

For light excited leptons ($m^\star \lesssim 200$ GeV), the attainable
limits at the LHC are less stringent than the bounds originating from
LEP \cite{lep}, but they are comparable to the limits obtained by
HERA \cite{hera}.  Notwithstanding, the LHC bounds are much
stronger than the presently available ones for a large range of
excited lepton masses (up to 2 TeV), being at least one order of
magnitude better. Furthermore, assuming $f=f^\prime$ and $f/\Lambda =
1/ m^\star$, the LHC will be able to exclude the existence of excited
leptons with masses up to 1 TeV.
A similar sensibility can be reached at the Next Linear Collider
(NLC) \cite{emg-mcgg-sfn}.

\medskip

In our analysis, we assumed that the excited leptons interact with the SM
particles via the effective operator (\ref{lag_1}). This is a conservative
assumption since it is possible that excited fermions may also couple to
ordinary quarks and leptons via contact interactions originating from the
strong constituent dynamics. In this case, the production cross section should
be enhanced \cite{quartic}. However, the contact interactions also modify 
the Drell-Yan process and can be strongly constrained if no deviation from 
the SM predictions is observed in this process.


\acknowledgments

This work was supported by Conselho Nacional de Desenvolvimento
Cient\'{\i}fico e Tecnol\'ogico (CNPq), by Funda\c{c}\~ao de Amparo \`a
Pesquisa do Estado de S\~ao Paulo (FAPESP), and by Programa de Apoio a
N\'ucleos de Excel\^encia (PRONEX).


\newpage 

\widetext

\begin{table}
\begin{tabular}{||c||c|c|c||c|c|c||}
$m^\star$(GeV) & $e^* \to e \gamma$ & $e^* \to e Z$ & $e^* \to \nu W$ & 
$\nu^* \to \nu \gamma$ & $\nu^* \to \nu Z$ & $\nu^* \to e W$\\
\hline
\hline
100  & 0.728 (0.) & 0.012 (0.137) & 0.260 (0.863) & 
       0. (0.728) & 0.137 (0.012) & 0.863 (0.260) \\
250  & 0.317 (0.) & 0.103 (0.381) & 0.580 (0.619) & 
       0. (0.317) & 0.381 (0.103) & 0.619 (0.580) \\
500  & 0.289 (0.) & 0.111 (0.391) & 0.600 (0.609) & 
       0. (0.289) & 0.391 (0.111) & 0.609 (0.600) \\
750  & 0.284 (0.) & 0.113 (0.393) & 0.603 (0.607) & 
       0. (0.284) & 0.393 (0.113) & 0.607 (0.603) \\
1000 & 0.282 (0.) & 0.113 (0.393) & 0.605 (0.607) & 
       0. (0.282) & 0.393 (0.113) & 0.607 (0.605)
\end{tabular}
\medskip
\caption{
 Branching ratios of excited leptons with the coupling constant
 assignment $f=f'\neq0$ ($f=-f'\neq0$). Notice that the branching
 ratios do not dependent on the value of $\Lambda$.}
\label{br}
\end{table}


\begin{table}
\begin{tabular}{||c||c||c||c||}
$\overline{M}$ (GeV) & 
$pp \to e^+ e^- \gamma$ & $pp \to e^+ e^-$ jet &
total \\
\hline
\hline
100  & 31.43   & 0.18    & 31.61   \\
250  & 3.55    & 0.04    & 3.59    \\
500  & 0.279   & 0.004   & 0.283   \\
750  & 0.051   & 0.001   & 0.052   \\
1000 & 0.0139  & 0.0002  & 0.0141  \\
1250 & 0.0048  & 0.0001  & 0.0049  \\
1500 - 2500 & 0.0243 & 0.0004 &  0.0247  
\end{tabular}
\medskip
\caption{ Cross sections in fb of the irreducible and jet faking photon
        backgrounds for the process $pp \to e^+ e^- \gamma$ after cuts
        (\ref{acceptance}) and (\ref{inv_mass_ee}).  For $\overline{M}$ up to
        1500 GeV, the invariant mass cut (\ref{inv_mass_eg}) was also imposed
        while for higher $\overline{M}$ the invariant mass cut
        (\ref{inv_mass_eg_hm}) was applied.  }
\label{back1}
\end{table}


\begin{table}
\begin{tabular}{||c||c||c||c||}
$\overline{M}$ (GeV) & 
$pp \to e^\pm \nu_{e^\pm} \gamma$ & 
$pp \to e^\pm \nu_{e^\pm}$ jet &
total \\
\hline
\hline
100  & 768.0   & 53.9  & 821.9  \\
250  & 51.65   & 2.94  & 54.59  \\
500  &  3.54   & 0.17  &  3.71  \\
750  &  0.74   & 0.03  &  0.77  \\
1000 &  0.23   & 0.01  &  0.24  \\
1250 &  0.087  & 0.002 & 0.089  \\
1500 - 2500 & 0.54   & 0.01 & 0.55 
\end{tabular}
\medskip
\caption{ Cross sections in fb of the irreducible and jet faking photon
        backgrounds for the process $pp \to e^\pm \nu_{e^\pm} \gamma$ after
        the cuts (\ref{acceptance}). For $\overline{M}$ up to 1500 GeV, the
        invariant mass cut (\ref{inv_mass_eg}) was also applied while for
        higher $\overline{M}$ the invariant mass cut (\ref{inv_mass_eg_hm})
        was imposed.}
\label{back2}
\end{table}


\begin{table}
\begin{tabular}{||c||c||c||c||}
$\overline{M}$ (GeV) & 
$pp \to e^\pm \nu_{e^\pm} \gamma$ & 
$pp \to e^\pm \nu_{e^\pm}$ jet &
total \\
\hline
\hline
100  & 857.3   & 68.3   & 925.6  \\
250  & 12.25   & 2.20   & 14.45  \\
500  &  0.57   & 0.13   &  0.70  \\
750  &  0.08   & 0.02   &  0.10  \\
1000 &  0.017  & 0.005  & 0.022  \\
1250 &  0.004  & 0.001  & 0.005  \\
1500 - 2500 & 0.029  & 0.007  & 0.036  
\end{tabular}
\medskip
\caption{ Cross sections in fb of the irreducible and jet faking photon
        backgrounds for the process $pp \to e^\pm \nu_{e^\pm} \gamma$ after
        the cuts (\ref{acceptance}). For $\overline{M}$ up to 1500 GeV, the
        invariant mass cut (\ref{trans_mass_ng}) was also used, while for
        higher $\overline{M}$ the invariant mass cut (\ref{trans_mass_ng_hm})
        was applied.}
\label{back3}
\end{table}


\begin{references}

 
\bibitem{experimental_data} 
D.~Charlton, ``Standard Model'', {\it Talk given at the International 
Europhysics Conference on High Energy Physics- HEP2001}
(http://www.hep2001.elte.hu/).

\bibitem{composite} For reviews and references see:\\
R.~S.~Chivukula, ``Lectures on technicolor and compositeness,''
arXiv:hep-ph/0011264; \\  
W.~Buchmuller, {\it Lectures given at 24th Int. Universitatswochen fur 
Kernphysik, Schladming, Austria, Feb 20 - Mar 1, 1985};\\
M.~E.~Peskin, {\it Presented to 1985 Int. Symp. on Lepton and Photon 
Interactions at High Energies, Kyoto, Japan, Aug 19-24, 1985};\\
M.~Suzuki, ``Survey of composite particle models of electroweak interaction,''
{\it Based on talk given at Int. Symp. on Bound Systems and Extended 
Objects, Karuizawa, Japan, Mar 19-21, 1992}.

\bibitem{excited_states}
F.~E.~Low, Phys.\ Rev.\ Lett.\  {\bf 14}, 238 (1965);\\
F.~Boudjema, Int.\ J.\ Mod.\ Phys.\ A {\bf 6}, 1 (1991);\\
M.~E.~Peskin, {\it Contributed to Proc. of 10th Int. Symp. on Lepton and 
Photon Interactions at High Energy, Bonn, West Germany, Aug 24-29, 1981}.

\bibitem{hera} 
C.~Adloff {\it et al.}  [H1 Collaboration],
Eur.\ Phys.\ J.\ C {\bf 17}, 567 (2000); \\
C.~Adloff {\it et al.}  [H1 Collaboration], arXiv:hep-ex/0110037; \\
S.~Chekanov {\it et al.}  [ZEUS Collaboration], arXiv:hep-ex/0109018.

\bibitem{lep} 
R.~Barate {\it et al.}  [ALEPH Collaboration],
Eur.\ Phys.\ J.\ C {\bf 4}, 571 (1998); \\
P.~Abreu {\it et al.}  [DELPHI Collaboration],
Eur.\ Phys.\ J.\ C {\bf 8}, 41 (1999); \\
M.~Acciarri {\it et al.}  [L3 Collaboration],
Phys.\ Lett.\ B {\bf 502}, 37 (2001); \\
G.~Abbiendi {\it et al.}  [OPAL Collaboration],
Eur.\ Phys.\ J.\ C {\bf 14}, 73 (2000).

\bibitem{chiral}
S.~J.~Brodsky and S.~D.~Drell, Phys.\ Rev.\ D {\bf 22}, 2236 (1980);
F.~M.~Renard, Phys.\ Lett.\ B {\bf 116}, 264 (1982).

\bibitem{effective_Lagrangian}
N.~Cabibbo, L.~Maiani, and Y.~Srivastava, 
Phys.\ Lett.\ B {\bf 139}, 459 (1984);\\
J.~Kuhn and P.~Zerwas, 
Phys.\ Lett.\ B {\bf 147}, 189 (1984);\\
K.~Hagiwara, D.~Zeppenfeld, and S.~Komamiya, 
Z.\ Phys.\ C {\bf 29}, 115 (1985);\\
F.~Boudjema and A.~Djouadi, 
Phys.\ Lett.\ B {\bf 240}, 485 (1990);\\
F.~Boudjema, A.~Djouadi, and J.~L.~Kneur, 
Z.\ Phys.\ C {\bf 57}, 425 (1993).

\bibitem{MadGraph}
T.~Stelzer and W.~F.~Long, Comput.\ Phys.\ Commun.\  {\bf 81}, 357 (1994).

\bibitem{helas}
H.~Murayama, I.~Watanabe, and K.~Hagiwara,
``HELAS: HELicity Amplitude Subroutines for Feynman diagram evaluations,''
KEK-91-11.

\bibitem{mrs} 
A.~D.~Martin, W.~J.~Stirling, and R.~G.~Roberts, Phys.\ Lett.\ 
{\bf B354}, 155 (1995).

\bibitem{pedro}
H.~Baer, P.~G.~Mercadante, F.~Paige, X.~Tata, and Y.~Wang,
Phys.\ Lett.\ B {\bf 435}, 109 (1998).

\bibitem{emg-mcgg-sfn} 
E.~M.~Gregores, M.~C.~Gonzalez-Garcia, and S.~F.~Novaes,
Phys.\ Rev.\ D {\bf 56}, 2920 (1997).

\bibitem{quartic} 
U.~Baur, M.~Spira, and P.~M.~Zerwas,
Phys.\ Rev.\ D {\bf 42}, 815 (1990); \\
 O.~\c {C}akir, C.~Leroy and R.~Mehdiyev, ATL-PHYS-2001-015.

\end{references}
\end{document}